\documentclass[12pt]{iopart}

\usepackage{bm}
\usepackage{graphicx, color}
\usepackage{wrapfig}
\usepackage{here}

\begin{document}
\title{Perturbative solution of a propagating interface in the phase
  field model}
\author{Mao Hiraizumi$^1$, Shin-ichi Sasa$^2$}
\address{Department of Physics, Kyoto University, Kyoto 606-8502, Japan}
\eads{hiraizumi.mao.72s@st.kyoto-u.ac.jp$^1$, sasa@scphys.kyoto-u.ac.jp$^2$}
\date{\today}
\if0
\fi
\begin{abstract}
When a stable ordered phase and a metastable disordered phase are separated by a flat interface, the metastable 
state changes to the stable state through the propagation of the interface.
For cases in which latent heat is generated, the interface displacement during some time interval is proportional
to the square root of the time interval when the extent of supercooling
is less than a certain value. We demonstrate this behavior by deriving
a perturbative solution for a propagating interface in the phase field
model. We calculate the leading-order contribution explicitly, and find
that the interface temperature deviates from the equilibrium transition
temperature in proportion to the interface velocity. 
\end{abstract}
%
% Uncomment for keywords
%\vspace{2pc}
%\noindent{\it Keywords}: XXXXXX, YYYYYYYY, ZZZZZZZZZ
%
% Uncomment for Submitted to journal title message
%\submitto{\JPA}
%
% Uncomment if a separate title page is required
%\maketitle
% 
% For two-column output uncomment the next line and choose [10pt] rather than [12pt] in the \documentclass declaration
%\ioptwocol
%
%%%%%%%%%%%%%%%%%%%%%%%%%%%%introduction
\section{Introduction}

Systems in which different phases coexist, such as growing crystal, exhibit a rich variety of
dynamical behaviors \cite{Langer,BG}.
The simplest situation is that a flat
interface connects a stable ordered phase and a metastable
disordered phase. In this case, the metastable state becomes 
stable through the propagation of the interface. The phenomenon
can be described by the time evolution of an order-parameter field
that represents the extent of the order. 

When the order parameter is the only relevant dynamical variable
of the system, the solution of the propagating interface is easily
determined \cite{Pomeau}.
However, as typically observed in crystallization, energy flow
becomes a significant physical quantity, because latent heat
is generated in the ordering process. In this case, a temperature
field also evolves under the influence of the generated latent heat, and this temperature field influences the
time evolution of the order parameter. The set of coupled equations is called the {\it phase field model}
\cite{fix,Cagphase,Langphase,CL,penrose,kobayashi}. Deriving a solution for the propagating interface
in  the phase field model is not a simple task.

Explicitly, let $T_s$ and $T_m$ be the temperatures of the heat baths
attached to the stable ordered phase in the left region and to the
metastable phase in the right region, respectively. Note that
$T_s$ and $T_m$ are less than the equilibrium transition temperature
$T_c$. The latent heat per unit volume is denoted by $L$ and the
specific heat at constant pressure per unit volume is denoted by $c_p$. 
An important dimensionless quantity is
\begin{eqnarray}
\Delta\equiv\frac{c_p(T_{c}-T_m)}{L},
\label{Deltadef}
\end{eqnarray}
which represents the extent of the supercooling. The interface position $R(t)$ at time $t$ depends
on $\Delta$. Specifically, numerical simulations of the phase
field model show that $R(t)\simeq t^{1/2}$ for the case $\Delta <1 $,
while $R(t)\simeq t$ for the case $\Delta > 1 $ \cite{Lowen}.

A theoretical problem is to derive this numerical observation
by analyzing the phase field model. For the case  $\Delta > 1$,
there is a solution describing 
$R(t)\simeq t $ \cite{CagNis}. We note that $R(t)\simeq t^{1/2}$ was derived
for the case $\Delta <1$ in the Stefan model,  which
formulates a dynamic boundary value problem \cite{Dewynne}.  
However, to the best of our knowledge, there have been  no theoretical
studies of the case $\Delta <1$ in the phase field model.

In this paper, we derive a perturbative solution of the propagating interface
for the case $\Delta <1$ in the phase field model.
A key step in deriving the solution is that
the solution is assumed to have a scaling form, with two scaled coordinates
and one dimensionless time-dependent small parameter. Expanding
the solution in the small parameter, we determine the leading-order
and next-order contributions to the solution. In particular, we obtain
\begin{eqnarray}
R(t)=\sqrt{2DCt},
\label{t1/2}
\end{eqnarray}
where $R(0)=0$, $D$ is a thermal diffusion constant, and 
$C$ is a constant determined by the boundary conditions of the
temperature field. As a remarkable property, $C$ exhibits
divergent behavior as $\Delta \to 1$ from below. The solution
also shows that the interface temperature
deviates from the equilibrium transition temperature in proportion
to the interface velocity.

%%%%%%%%%%%%%%%%%%%%%%%%%%%%%%%%
\section{Phase field model}

% equation of $\phi$
Since we are focusing on the motion of a flat interface in a three-dimensional
space, we study a one-dimensional system.
Let $\phi(x,t)$ be an order-parameter field that represents
the ordered (e.g., solid) phase by $\phi(x,t)=1$ and the disordered
(e.g., liquid) phase by $\phi=0$.
The phase field model is a set of coupled equations for
the order-parameter field
$\phi(x,t)$ and the temperature field $T(x,t)$. We first assume
a free energy density $f(\phi,T)$ for a given material, where 
$f(\phi,T)$ takes a double-well form that possesses local
minima at $\phi = 0$ and $\phi=1$ for each $T$. The transition
temperature $T_c$ is determined such that $f(\phi,T)$
is minimized at $\phi=1$ when $T < T_c$, while $f(\phi,T)$
is minimized at $\phi=0$ when $T > T_c$.
We then define the free energy functional
$F[\phi,T]$ as 
\begin{equation}
F[\phi,T]=\int dx \left(f(\phi,T)+\frac{\xi^2}{2}(\partial_x\phi)^2\right),
\end{equation}
where $\xi$ is a parameter that represents the interface width.
The gradient term becomes relevant in the interface region, 
corresponding to the surface free energy. We assume that $\phi$ evolves
so that the free energy decreases. That is, the equation of  $\phi$
is given as
\begin{equation}
\tau\partial_t \phi=-\frac{\delta F[\phi,T]}{\delta \phi(x)}
\label{phi_eq_1},
\end{equation}
where $\tau$ characterizes the time scale of $\phi$.

% equation of $T$
From the law of enthalpy conservation at constant pressure, the
equation for $T$ is determined as 
\begin{eqnarray}
  \partial_t T = D\partial_x^2 T
  +\frac{L}{c_p}\frac{\partial \phi}{\partial t}
\label{T_eq},
\end{eqnarray}
where  the thermal diffusion constant $D$ is assumed to be
independent of $(T,\phi)$.
The first term on the right-hand side of (\ref{T_eq}) 
represents  heat diffusion and the second term represents
the effect of the latent heat generated by the time evolution of $\phi$. 
To simplify the notation, we introduce
a dimensionless temperature
\begin{equation}
\theta(x,t)\equiv  \frac{c_p(T(x,t)-T_c)}{L},
\end{equation}
and define
\begin{equation}
  \bar f(\phi,\theta)\equiv f\left( \phi, T_c+\frac{L}{c_p} \theta \right).
\end{equation}
Using these quantities, we can rewrite (\ref{phi_eq_1}) and (\ref{T_eq}) as 
\begin{eqnarray}
  \tau\partial_t \phi &= &\xi^2 \partial_x^2 \phi
  -\frac{\partial \bar f}{\partial \phi},
\label{phi_eq_2}
  \\
  \partial_t \theta & = &D\partial_x^2 \theta+\partial_t \phi.
\label{u-eq}  
\end{eqnarray}

% Boundary condition

We study cases in which a stable ordered phase (e.g., crystal) in the left-region
grows in a metastable disordered phase (e.g., supercooled liquid)
in the right
region. Thus, we impose the following boundary
conditions on $\phi$ and $\theta$:   
\begin{eqnarray}
  \lim_{x \to -\infty} \phi(x,t) &=& 1  , \label{bc:1}\\
  \lim_{x \to \infty} \phi(x,t) &=& 0 , \label{bc:0}
\end{eqnarray}
and 
\begin{eqnarray}
  \lim_{x \to -\infty} \theta(x,t) &=& -\Pi   \label{bc:Pi}, \\
  \lim_{x \to \infty} \theta(x,t) &=& -\Delta \label{bc:Delta},
\end{eqnarray}
where $\Delta$ is defined as (\ref{Deltadef}) and  $\Pi$ is 
\begin{equation}
  \Pi \equiv - \frac{c_p(T_{s}-T_c)}{L}.
\end{equation}
Note that $\Pi \ge 0$. More precisely, we assume that $\phi(x,t)$ and $\theta(x,t)$
converge faster than an exponential form as a function of $x$ in the
limit $|x| \to \infty$.

\section{Results}

% setup

We construct a solution of (\ref{phi_eq_2}) and (\ref{u-eq})
with the boundary conditions (\ref{bc:1}), (\ref{bc:0}), (\ref{bc:Pi}),
and (\ref{bc:Delta}).
Let $R(t)$ be the interface position at time $t$, which satisfies
$\phi(R(t),t)=0.5$. First, we note that $\xi$
and  $D/\dot{R}$ provide length scales of the solution $\phi(x,t)$ and
$\theta(x,t)$, respectively, where $\dot R (t)=dR(t)/dt$.
Based on this observation, we define the
scaled coordinates
\begin{eqnarray}
w &\equiv& \frac{x-R(t)}{\xi}, \\
z &\equiv& \frac{\dot{R}(x-R(t))}{D}.
\end{eqnarray}
We also introduce a small dimensionless time-dependent quantity 
\begin{equation}
\eta \equiv\frac{\xi \dot{R}}{D}.
\end{equation}
Using $w$, $z$, and $\eta$, we  assume that the
solution takes the scaling form
\begin{eqnarray}
\phi(x,t) &=& \Phi(w;\eta),  \\
\theta(x,t) &=& \Theta(z;\eta).
\end{eqnarray}
Focusing on cases where $\eta(t) \ll 1 $, we expand the solution
in $\eta$ as 
\begin{eqnarray}
\Phi(w;\eta)&=&\Phi_0(w)+\eta \Phi_1(w)+O(\eta^2) , \label{a}\\
\Theta(z;\eta)&=&\Theta_0(z)+\eta \Theta_1(z)+ O(\eta^2) .\label{b}
\end{eqnarray}

% equation for \Phi

Substituting (\ref{a}) and (\ref{b}) into (\ref{phi_eq_2})
and extracting terms that are independent of $\eta$, we have
\begin{equation}
  \partial_w^2 \Phi_0-\frac{\partial \bar f}{\partial \phi} (\Phi_0,\Theta_0(0)) 
  =0,  \label{Phi00}
\end{equation}
where we have used $z=\eta w$. Multiplying both sides of
(\ref{Phi00}) by $\partial_w \Phi_0$ and integrating the result from $-\infty$ to $+\infty$, we obtain 
\begin{eqnarray}
\bar f(1,\Theta_0(0))=\bar f(0,\Theta_0(0)), \label{U0=0}
\end{eqnarray}
which leads to
\begin{equation}
  \Theta_0(0)=0.
\label{U0-0-0}
\end{equation}
By solving ($\ref{Phi00}$) with the boundary conditions (\ref{bc:1}) and (\ref{bc:0}), we can
determine the unique solution $\Phi_0(z)$.

% equation for  U

Next, substituting (\ref{a}) and (\ref{b}) into (\ref{u-eq})
and extracting the leading-order terms in $\eta$, 
we obtain
\begin{equation}
  \left(\frac{\ddot{R }}{\dot{R}}z-\frac{\dot{R}^2}{D} \right)
  \partial_z\Theta_0 
  =\frac{\dot{R}^2}{D}\partial_z^2 \Theta_0
  + \frac{\dot{R}^2}{D} \delta(z),
\label{Ueq}
\end{equation}
where $\ddot{R }=d^2 R(t)/dt^2$; here, we have used 
\begin{equation}
\lim_{\eta \to 0}
\frac{1}{\eta}\partial_w\Phi_0(w)|_{w=\frac{z}{\eta }}=-\delta(z).
\end{equation}
%because $\Phi_0(z/\eta)\to 1-\theta(z)$ in the limit $\eta \to 0$.
By the method of separation of variables, we find
\begin{eqnarray}
  \frac{D\ddot{R}}{\dot{R}^3}=-\frac{1}{C}
\label{RRdot}
\end{eqnarray}
with a constant $C$.  From this relation, we obtain (\ref{t1/2}).  
Then, (\ref{Ueq}) becomes
\begin{eqnarray}
\partial_z^2 \Theta_0+\frac{1}{C}(z+C)\partial_z\Theta_0=-\delta(z). \label{U0} 
\end{eqnarray}

% solution of U_0

\begin{figure}[H]
\begin{minipage}[t]{0.5\hsize}
\centering
\includegraphics[width=60mm]{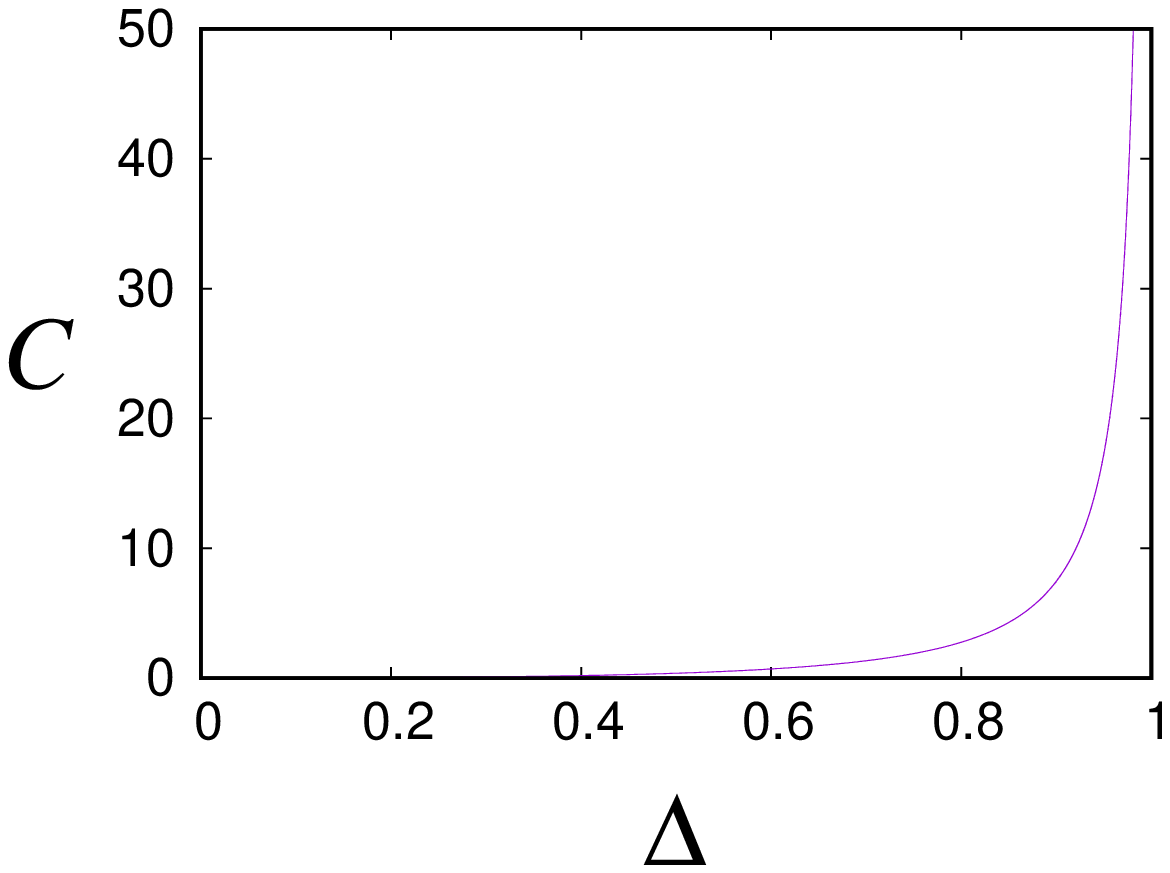}
\hspace{-10mm}
\end{minipage}
\begin{minipage}[t]{0.5\hsize}
\centering
\includegraphics[width=60mm]{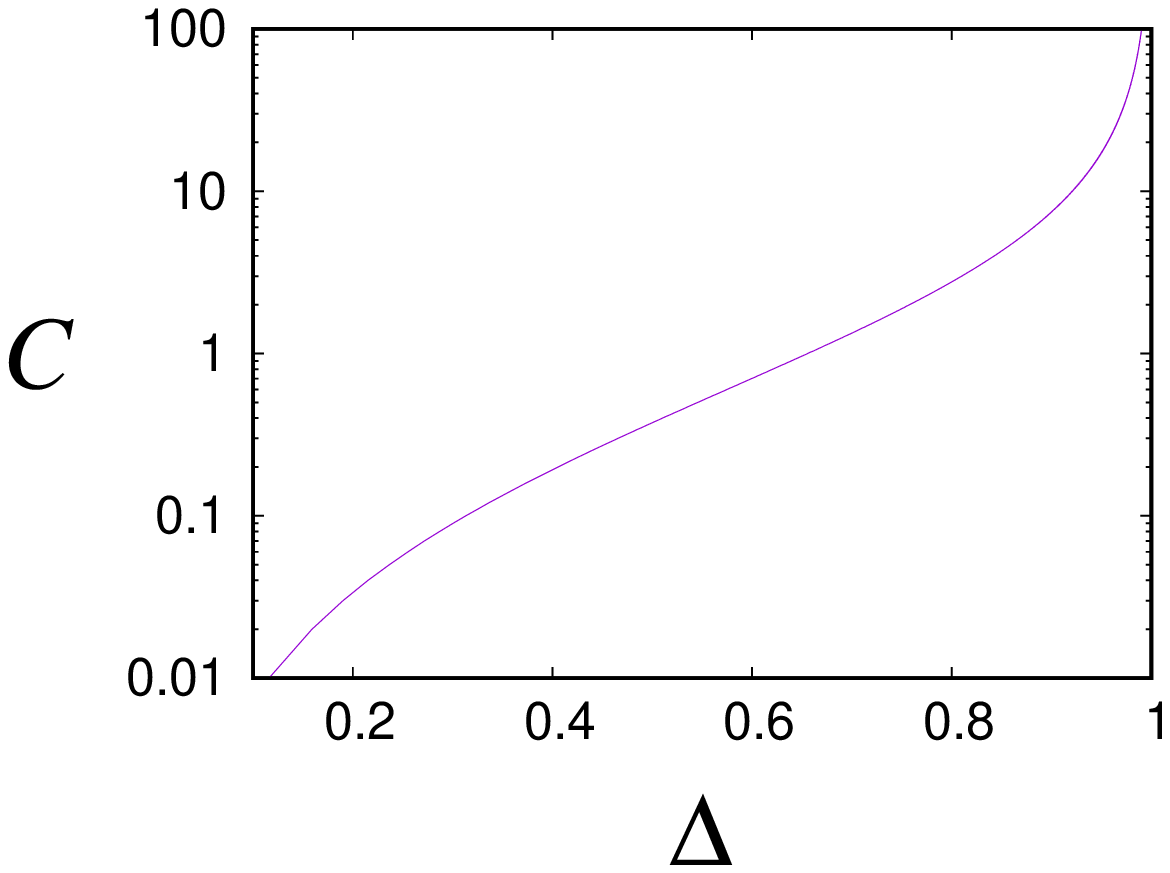}
\hspace{-10mm}
\end{minipage}
\caption{$C$ as a function of $\Delta$ for the case $\Pi=0$.
Normal and log plots are  displayed on the left 
and right sides, respectively. $C \to \infty$ as $\Delta \to 1$. }
\label{fig1}
\end{figure}

We now analyze (\ref{U0}). First, set $\Theta^{+}(z)=\Theta(z)$ for $z >0$ and 
$\Theta^{-}(z)=\Theta(z)$ for $z <0$.  Solving (\ref{U0}), we obtain
\begin{eqnarray}
  \Theta_0^{\pm}(z)&=&\partial_z \Theta_0|_{z=\pm 0}
  e^{\frac{C}{2}}\int_0^z e^{ -\frac{(z^\prime+C)^2}{2C}  }dz^\prime\\
  &=&\partial_z\Theta_0|_{z=\pm 0}\sqrt{\frac{\pi C}{2}}
  e^{\frac{C}{2}}\left(\rm{erf}
  \left(\frac{z}{\sqrt{2C}}
  +\sqrt{\frac{C}{2}} \right)-\rm{erf}\left(\sqrt{\frac{C}{2}}\right)\right)
\end{eqnarray}
with the connection condition
\begin{eqnarray}
  &\partial_z\Theta^+_0|_{z=+0}-\partial_z \Theta^-_0|_{z=-0}=-1.
\label{BC}
\end{eqnarray}
For later convenience, we set
\begin{equation}
  A \equiv\partial_z\Theta^-|_{z=-0} .
\label{Adef}
\end{equation}
Using the boundary conditions (\ref{bc:Pi}) and (\ref{bc:Delta}), 
we derive
\begin{eqnarray}
  \Delta&=& (1-A) \sqrt{\frac{\pi C}{2}}e^{\frac{C}{2}}
  \left(1-\rm{erf}
  \left(\sqrt{\frac{C}{2}} \right)
  \right ),
  \label{Deltaeq} \\
  \Pi&= &A \sqrt{\frac{\pi C}{2}}e^{\frac{C}{2}}
  \left(1+\rm{erf}\left(\sqrt{\frac{C}{2}} \right) \right).
  \label{Pieq}
\end{eqnarray}
Therefore,  for a given value of $(\Delta, \Pi)$ satisfying $\Delta <1$,
we can determine  $A$ and $C$. 
As one example, we consider the case $\Pi=0$. In this case, we
have $A=0$ from (\ref{Pieq}), and  we obtain $C$ as a function of
$\Delta$. The result is displayed in Fig.~\ref{fig1}. Note
that $C \to \infty$ as $\Delta \to 1$. 

% higher order contribution

Furthermore, we substitute (\ref{a}) and (\ref{b}) into (\ref{phi_eq_2})
and (\ref{u-eq}) and collect terms proportional to $\eta$, where 
we also assume the expansion 
\begin{eqnarray}
  \frac{D\ddot{R}}{\dot{R}^3}=-\frac{1}{C}-C_1\eta+O(\eta^2).
\label{RRdot2}
\end{eqnarray}
We then obtain equations for $\Phi_1(w)$ and $\Theta_1(z)$. By solving these
equations with the boundary conditions that
$\Phi_1(w)$ and $\Theta_1(z)$ converge to zero faster than an
exponential form in the limit $|w| \to \infty$ and $|z| \to \infty$,
we can determine $\Phi_1(w)$, $\Theta_1(z)$, and $C_1$.  (See \ref{app-sec1}.)
For example, the interface temperature $T(R(t),t)$
is determined as
\begin{equation}
  T(R(t),t)=T_c+\frac{L\xi \dot R(t)}{c_pD} \Theta_1(0)
\end{equation}
with the formula for $\Theta_1(0)$ given in (\ref{alpha}).
This means that the interface temperature deviates from the equilibrium
transition temperature and the deviation is proportional to the interface
velocity $\dot R$.

\section{Concluding remarks}

% noise
We have studied the deterministic equation describing phase coexistence.
However, because the interface length $\xi$ is less than a micrometer,
stochastic processes may become relevant in the interface region. To consider such effects, we examine an energy-conserving Potts model with
kinetic energy as a natural extension of energy-conserving kinetic Ising
models \cite{KS,Creutz,Casartelli,HOS}. We will report the behavior in a separate
paper.

% instability 

In this paper, we focused on the propagation of a flat interface as the
simplest case. In some experiments, flat interfaces have exhibited the
Mullins--Sekerka instability \cite{MS}, which leads to a rich
variety of patterns \cite{Langer}. The instability of propagating
interfaces was studied in the phase field model \cite{kupf, braun},
including the instability of solutions with $R(t)\simeq t^{1/2} $
\cite{Lambert}. With regard to the propagation
velocity of a destabilized interface, it was reported that $R(t)\simeq t $
when $\Delta >1$ in the one-sided model \cite{JV}, while a theoretical analysis
of the Stefan-type model shows $R(t)\simeq t^{1/2}$ even when
$\Delta>1$ \cite{Cag}. A systematic understanding of these phenomena
will be studied in the future. 
\ack

%% acknowledgements

This work was supported by KAKENHI (Grant Nos. 17H01148, 19H05795, and 20K20425).

%%%%%%%%%%%%%appendix
\appendix
\section{Next-order contribution}
\label{app-sec1}
In this appendix, we calculate the next-order contribution to the solution. To simplify the notation, we introduce a dimensionless parameter
\begin{equation}
  \alpha \equiv \frac{D\tau}{\xi^2},
\end{equation}
which represents the ratio of the thermal diffusion constant
to the order-parameter diffusion constant.

%%%%
First, we substitute (\ref{a}) and (\ref{b}) into (\ref{phi_eq_2}).
We collect terms proportional to $\eta$, noting that
$\dot \eta =O(\eta^3)$ from (\ref{RRdot}).
We then obtain
\begin{eqnarray}
  -\alpha \partial_w\Phi_0(w)&=& \partial_w^2 \Phi_1
  -\Phi_1 \frac{\partial^2 \bar f}{\partial \phi^2}(\Phi_0,0)
  \nonumber \\
&-  &\frac{\partial^2 \bar f}{\partial \phi \partial \theta}(\Phi_0,0)
\Bigl( w \Theta_0^\prime(\pm 0)  +  \Theta_1(0)\Bigr), \label{Phi11}
\end{eqnarray}
where either $+0$ or $-0$ is selected in the second line depending
on whether $w >0$ or $w<0$.
Multiplying both sides of (\ref{Phi11}) by $\partial_w \Phi_0$ and
integrating the result from $-\infty$ to $\infty$,  we obtain 
\begin{equation}
  -\alpha I=- \Theta_1(0) J+(1-A)K_+  - A K_-   \label{alpha}
\end{equation}
with 
\begin{eqnarray}
  I &\equiv& \int_{-\infty}^\infty dw (\partial_w\Phi(w))^2, \\
  J &\equiv& \int_{-\infty}^\infty dw
  \frac{\partial^2 \bar f}{\partial \phi \partial \theta}(\Phi,0) \partial_w\Phi, \\
  K_{\pm} &\equiv & \pm \int_{0}^{\pm \infty} dw
  w\frac{\partial^2 \bar f}{\partial \phi \partial \theta}(\Phi,0)
  \partial_w\Phi .
\end{eqnarray}
Here, we have also used (\ref{Adef}). 
Equation (\ref{alpha}) determines the value of $\Theta_1(0)$,
and corresponds to  the solvability condition for the equation 
of $\Phi_1(w)$. Then, $\Phi_1(w)$ can be obtained as the solution
of the linear equation (\ref{Phi11}).

% equation for  U

Next, we substitute (\ref{a}) and (\ref{b}) into (\ref{u-eq}).
Collecting terms proportional to $\eta$, we obtain
\begin{equation}
  \partial_z^2 \Theta_1 +\partial_z\Theta_1
  +\frac{1}{C}\partial_z(z\Theta_1)=C_1z\partial_z \Theta_0 .
  \label{U1} 
\end{equation}
By integrating (\ref{U1}) from $-\infty$ to $+\infty$, we have 
\begin{equation}
C_1=0,
\end{equation}
which corresponds to the solvability condition for the equation
of $\Theta_1(z)$. Then, (\ref{U1}) becomes
\begin{eqnarray}
\partial_z \Theta_1+\frac{1}{C}(z+C)\Theta_1=B,
\label{U2} 
\end{eqnarray}
where $B$ is a constant. Because $\Theta_1(z)$ converges to zero
faster than an exponential form as a function of $|z|$ in the
limit $|z| \to \infty$, we find that
\begin{equation}
B=0.
\end{equation}
We then solve (\ref{U2}) as 
\begin{equation}
\Theta_1(z)=\Theta_1(0)e^{\frac{C}{2}}e^{-\frac{(z+C)^2}{2C}},
\end{equation}
where $\Theta_1(0)$ is given by ($\ref{alpha}$).
 
%% references
\section*{References}
%\vspace{10mm}
 
\end{document}